# "Groupware for Groups":
# Problem-Driven Design in *Deme*

**Todd Davies**
Symbolic Systems Program
Stanford University, Stanford, CA, 94305-2150
tdavies@csli.stanford.edu

**Alex Cochran**
Symbolic Systems Program
Stanford University, Stanford, CA 94305-2150
alex.cochran@stanford.edu

**Brendan O'Connor**
Symbolic Systems Program
Stanford University, Stanford, CA 94305-2150
brendano@stanford.edu

**Andrew Parker**
Symbolic Systems Program
Stanford University, Stanford, CA 94305-2150
aparker@stanford.edu

**ABSTRACT**
Design choices can be clarified when group interaction software is directed at solving the interaction needs of particular groups that pre-date the groupware. We describe an example: the Deme platform for online deliberation. Traditional threaded conversation systems are insufficient for solving the problem at which Deme is aimed, namely, that the democratic process in grassroots community groups is undermined both by the limited availability of group members for face-to-face meetings and by constraints on the use of information in real-time interactions. We describe and motivate design elements, either implemented or planned for Deme, that addresses this problem. We believe that "problem focused" design of software for pre-existing groups provides a useful framework for evaluating the appropriateness of design elements in groupware generally.

**Author Keywords**
Online deliberation, groupware, social software, community networks.

**ACM Classification Keywords**
H5.3. Information interfaces and presentation: Group and organization interfaces – *asynchronous interaction, web-based interaction, computer-supported cooperative work*.

**INTRODUCTION**
The Partnership for Internet Equity and Community Engagement (PIECE) [1], a joint project of the East Palo Alto Community Network and the Symbolic Systems Program at Stanford University, has recently led to the creation of a platform for online deliberation called *Deme* (which rhymes with "team") [2]. Deme is designed to allow groups of people to engage in collaborative drafting, focused discussion, and decision making using the web.

In many ways Deme replicates the functionality of other groupware, being built around "meeting areas" within each "group space". Each meeting area includes a feature for threaded discussion, but also a number of other, less traditional features. Deme is designed for small to medium-sized groups that (a) have a substantial face-to-face existence that pre-dates or is independent of any interaction on the Internet, (b) are geographically limited so that all members can meet each other face to face, and (c) have difficulty meeting face-to-face as much as they need or would like to. Examples of such groups include neighborhood associations, places of worship, community interest groups, university groups (e.g. dormitories), and coalitions of activists.

**FROM PROBLEM TO GOAL**
In an earlier paper [3], members of our team documented a number of difficulties for community democracy in East Palo Alto that we attributed to a mismatch between face-to-face meetings and the reality of residents' lives there. As a byproduct of the technology boom in Silicon Valley, large amounts of money became available for both private and public initiatives in low-income, multi-ethnic and multi-lingual East Palo Alto early in this decade. But several factors – including diverse work schedules, long commutes, family demands, lack of key resources and information (including a lack of community media), historical mistrust of those in power, and cultural and language barriers – made it difficult for community and organization meetings to achieve sufficient participation for attaining full legitimacy. Although much was achieved, lasting rifts were created and significant opportunities were missed. Moreover, the necessity of making decisions synchronously and face-to-face, as elsewhere, provided an excuse for inner-circle decision making, as not everyone could be present to provide input.

Our participant-observation of groups in East Palo Alto led us to conclude that we were seeing instances of a pattern common to many groups in the contemporary United States. Other authors have noted the declining participation of citizens in civil society [4,5]. But East Palo Alto's new community network project [6], and the high interest and/or ability we found among most residents in using email and the web, convinced us that the Internet could facilitate greater community engagement in East Palo Alto and in many other communities. Many barriers existed, including various digital divides that are also a focus of our work. But there was one barrier we felt called for a more general solution: existing groupware systems (listservs and message boards) that were available to volunteer groups of citizens (who lack sufficient money to spend on intranet or meeting software) do not generally have the functionality to allow such groups to conduct many important business functions asynchronously, even given that residents could access and use the web with moderate proficiency.

A bit of preliminary thought about what is possible with present technology led to the formulation of our goal: creating a web platform where groups could asynchronously accomplish deliberative work that would otherwise require face-to-face meetings.

**FROM GOAL TO DESIGN**
In a longer paper [7], we have detailed the principles we want Deme to embody, fleshing out the goal of building a meeting-worthy asynchronous platform. Here we focus just on the most important design features that are motivated by *one* of the principles: *comprehensive* support for online, asynchronous performance of face-to-face meeting tasks.

**(1) Item-focused discussion.** Although global comments are allowed, Deme encourages posting and commenting on *items*, which function like agenda items in a meeting. "Item references" appear in comment headers, and display the item when clicked. An item can be a document, another website, a poll or decision, a question, or a project plan. Each of these is a separate item type, and more types may be created in the future. In the standard view of a meeting area (Figure 1), the screen is split so that items (an index or a single item) are displayed on the left and the discussion is displayed on the right. We believe this attempt to mimic the structure of meetings is a novel feature of Deme.

**(2) Flexible polls and decisions.** Nonbinding polls (not yet implemented) and decisions/votes can be created very flexibly, with multiple options for polling methods and decision rules so that groups may use the method they are accustomed to or which is in their bylaws. Votes can be changed up to a chosen deadline, and can be annotated.

**(3) In-text comments in documents.** Another feature that appears to be novel in Deme is that document items posted as text files allow comments to be inserted in any space. Comments appear in the discussion, linked via item references and a "comment reference" at the insertion space in the document. General comments can also be attached to documents, as to other item types.

**(4) Nondestructive document revision.** Users can edit a document and post the revision as an alternative. The current system does not allow wiki-style editing of documents because that requires further support for version recovery in the discussion if the integrity of comments is not to be lost, but this enhancement is planned.

**(5) Project management tool integrated with discussion.** Users can create "project" item types, similar to Bugzilla, with multiple editable and sortable tasks listing priority, handlers, status, etc. Users can volunteer for tasks. Projects and tasks can have comments attached to them.

**(6) Customizable group website.** The group homepage (Figure 2) provides access to standard group-level resources (group description, announcements, links, etc.), and Deme allows the group to be defined as either open or closed.

**(7) Optional sorting methods.** The discussion index threads can be viewed by subject, item, date, or author, and project task lists can be viewed in many different ways.

**(8) Expandable viewers.** The item and comment viewers can be expanded to fill the whole screen via a button at the upper right, popping up a new window.

**(9) Flexible integration with email.** Meeting areas currently send email notification to subscribed users whenever a comment or item is posted. Posting by email is in the process of being implemented. This is important both for pre-existing groups that already use an email list, and because many people use email as their primary messaging medium.

**(10) Multiple meeting areas per group space.** Deme attempts to mirror the structure of groups, which often have multiple committees and discussion groups internally.

**(11) Ability to share meeting areas across groups.** Not yet implemented, but an important feature for supporting networking across groups, and not commonly available.

**A GENERAL CLAIM**
Although an "early focus on users and tasks" has long been emphasized in the theory and practice of user-centered interaction design [8], groupware for the web often pre-dates or creates its community of users. The fact that software can create a group is enticing for a designer, but it does tend to lead to a "solution focused" rather than a "problem focused" approach [9]. We suspect, following Kruger and Cross's study of industrial designers [10], that problem-driven design is likely to lead to more effective groupware in many practical contexts.

In any case, a clear problem/goal focus provides a framework for evaluating different systems. We find that design decisions within our team are relatively easy to agree on once we have a common understanding of the problem we are trying to solve. Considering our target users led us not to include a reputation or rating system, for instance, because these appeared less useful for real-world groups and might intimidate less experienced web users. And considering groups in East Palo Alto helped us to determine what features are needed but missing in other systems for meeting-level use.

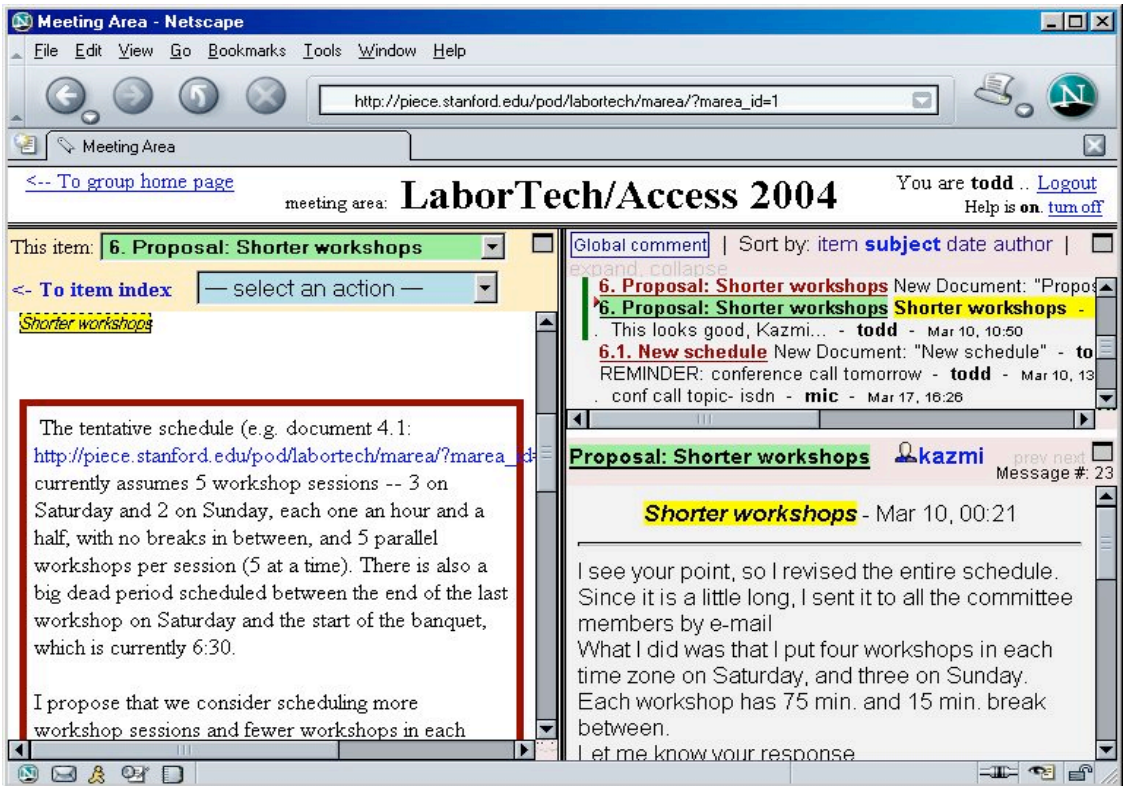

**Figure 1. Standard view of a meeting area**

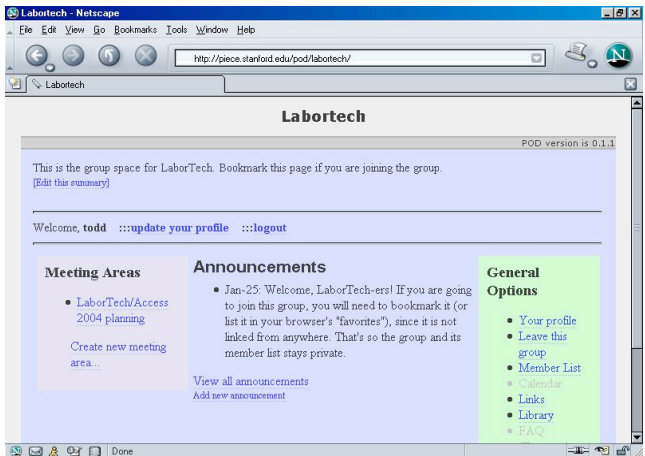

**Figure 2. A group homepage**

**ACKNOWLEDGMENTS**
See http://www.groupspace.org/development.php.

**REFERENCES**
1. PIECE website. http://piece.stanford.edu.
2. Deme website. http://www.groupspace.org.
3. Davies, T., Sywulka, B., Saffold, R., and Jhaveri, R. Community democracy online: A preliminary report from East Palo Alto. *APSA 2002 Online Proceedings*. http://apsaproceedings.cup.org/Site/abstracts/030/030006DaviesTodd.htm.
4. Putnam, R.D. *Bowling Alone: The Collapse and Revival of American Community*. Simon and Schuster (2000).
5. Skocpol, T. *Diminished Democracy: From Membership to Management in American Civic Life*. University of Oklahoma Press (2003).
6. EPA.Net. http://www.epa.net.
7. Davies, T., O'Connor, B., Cochran, A.A., Effrat, J.J. An online environment for democratic deliberation: motivations, principles, and design (2004). http://www.stanford.edu/~davies/deme-principles.pdf.
8. Gould, J.D. and Lewis, C.H. Designing for usability: key principles and what designers think. *Comm ACM, 28*, 3 (1985), 300-311. Quoted in Preece, J., Rogers, Y., and Sharp, H. *Interaction Design: Beyond Human-Computer Interaction*. John Wiley and Sons (2002), p. 285.
9. Lawson, B. Cognitive strategies in architectural design. *Ergonomics, 22*, 1 (1979), 59-68. Quoted in [10], p. 60.
10. Kruger, C. and Cross, N. Modelling cognitive strategies in creative design. In *Computational and Cognitive Models of Creative Design V* (2001). http://design.open.ac.uk/people/academics/cross/CogStratHI.pdf.